\documentclass[sigconf]{acmart}

\AtBeginDocument{%
  \providecommand\BibTeX{{%
    \normalfont B\kern-0.5em{\scshape i\kern-0.25em b}\kern-0.8em\TeX}}}

\usepackage{booktabs} 
\usepackage{float}
\usepackage{caption}
\usepackage{subcaption}

\usepackage{xspace}
\newcommand{\idest}{i.e.,\xspace}
\newcommand{\eg}{e.g.,\xspace}
\newcommand{\palpha}{\ensuremath{\text{P}^3\!\alpha}\xspace}
\newcommand{\palphabold}{\ensuremath{\text{\textbf{P}}^\textbf{3}\!{\boldsymbol \alpha}}\xspace}
\newcommand{\pbeta}{\ensuremath{\text{RP}^3\!\beta}\xspace}
\newcommand{\pbetabold}{\ensuremath{\text{\textbf{RP}}^\textbf{3}\!{\boldsymbol \beta}}\xspace}
\newcommand{\EASER}{EASE$^R$\xspace}

\setcopyright{rightsretained}

\begin{document}

\copyrightyear{2021} 
\acmYear{2021} 
\acmConference[UMAP '21 Adjunct]{Adjunct Proceedings of the 29th ACM Conference on User Modeling, Adaptation and Personalization}{June 21--25, 2021}{Utrecht, Netherlands}
\acmBooktitle{Adjunct Proceedings of the 29th ACM Conference on User Modeling, Adaptation and Personalization (UMAP '21 Adjunct), June 21--25, 2021, Utrecht, Netherlands}\acmDOI{10.1145/3450614.3461680}
\acmISBN{978-1-4503-8367-7/21/06}

\title{A Methodology for the Offline Evaluation of Recommender Systems in a User Interface with Multiple Carousels}

\author{Nicol\`o Felicioni}
\orcid{0000-0002-3555-7760}
\affiliation{%
  \institution{Politecnico di Milano, Italy}
}
\email{nicolo.felicioni@polimi.it}

\author{Maurizio Ferrari Dacrema}
\orcid{0000-0001-7103-2788}
\affiliation{%
  \institution{Politecnico di Milano, Italy}
}
\email{maurizio.ferrari@polimi.it}

\author{Paolo Cremonesi}
\orcid{0000-0002-1253-8081}
\affiliation{%
  \institution{Politecnico di Milano, Italy}
  }
\email{paolo.cremonesi@polimi.it}



\begin{abstract}
Many video-on-demand and music streaming services provide the user with a page consisting of several recommendation lists, \idest \emph{widgets} or \emph{swipeable carousels}, each built with a specific criterion (\eg most recent, TV series, etc.).
Finding efficient strategies to select which carousels to display is an active research topic of great industrial interest. 
In this setting, the overall quality of the recommendations of a new algorithm
cannot be assessed by measuring solely its individual recommendation quality. Rather, it
should be evaluated in a context where other recommendation lists are already available, to account for how they complement each other.  
This is not considered by traditional offline evaluation protocols. Hence, we propose an \emph{offline evaluation protocol for a carousel setting} in which the recommendation quality of a model is measured by how much it improves upon that of an already available set of carousels.
We report experiments on publicly available datasets on the movie
domain
and notice
that under a carousel setting the ranking of the algorithms change. 
In particular, when a SLIM carousel is available, 
matrix factorization models tend to be preferred, while item-based models
are penalized.
We also propose to extend ranking metrics to the two-dimensional carousel layout in order to account for a known position bias, \idest users will not explore the lists sequentially, but rather concentrate on the top-left corner of the screen.
\end{abstract}

\ccsdesc[500]{Information systems~Collaborative filtering}
\ccsdesc[500]{Information systems~Recommender systems}
\ccsdesc[500]{General and reference~Evaluation}

\keywords{Recommender Systems; User Interface; Evaluation}

\maketitle

\section{Introduction}
\begin{figure*}[]
    \centering
    \begin{minipage}{.45\textwidth}    
        \centering
        \includegraphics[width=1.\textwidth]{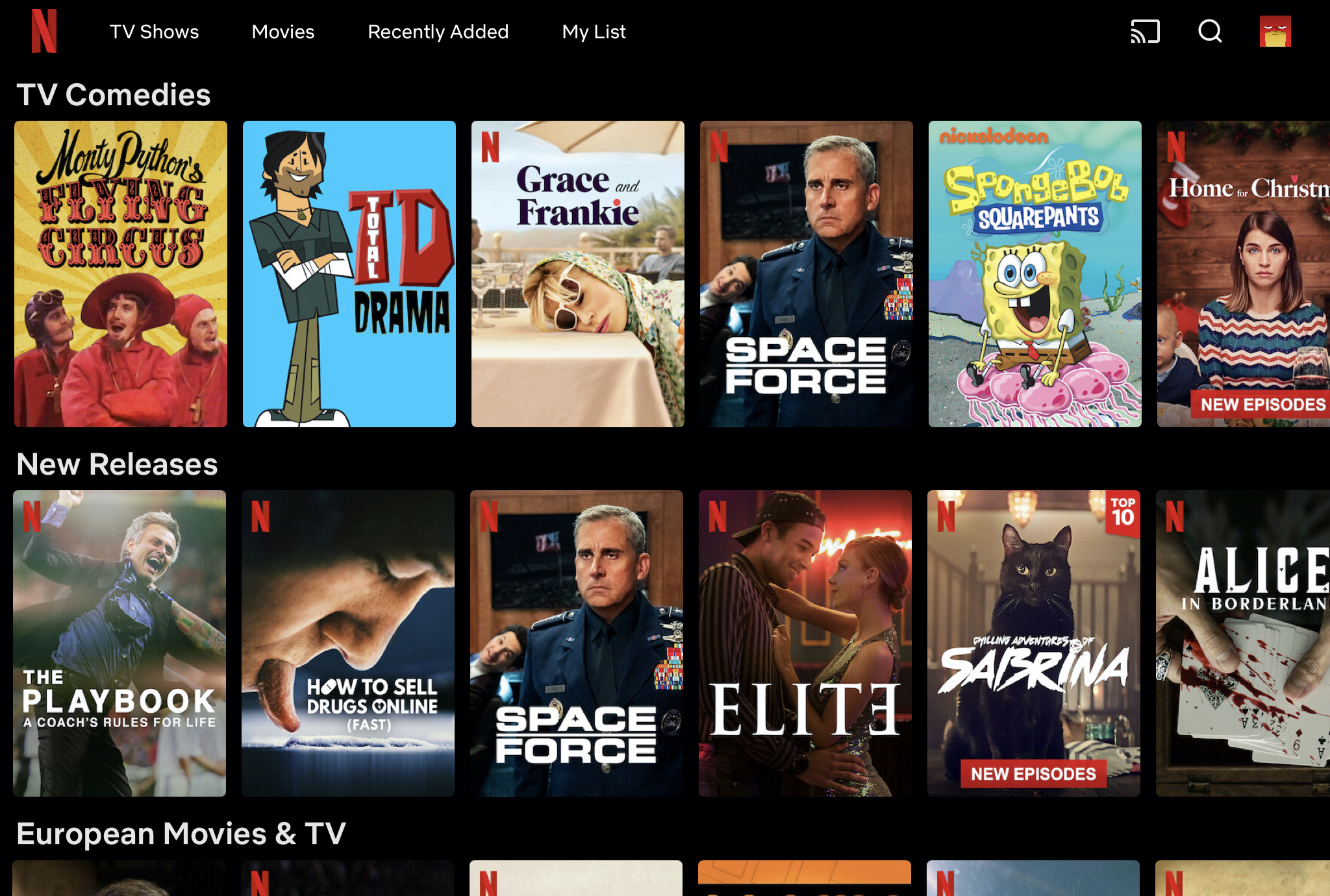}
        \caption{The Netflix homepage, an example of carousel user interface in the multimedia streaming domain.}
        \label{fig:amazon_netflix}
    \end{minipage}%
    \hfill
    \begin{minipage}{0.45\textwidth}
        \centering
        \includegraphics[width=1.\textwidth]{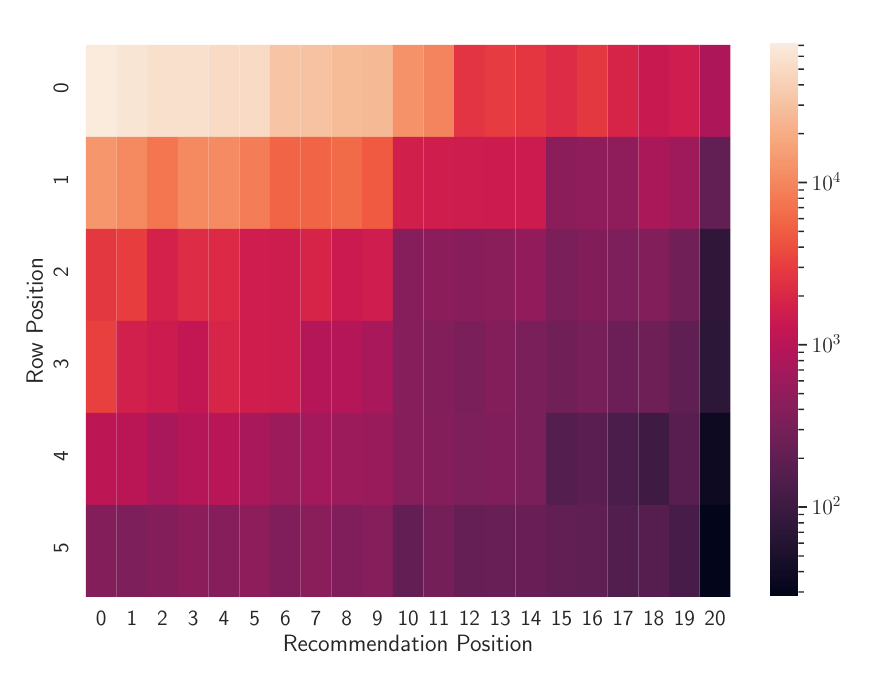}
        \caption{Heatmap of the number of interactions per position on the screen, taken from
        the dataset presented in 
        \cite{DBLP:conf/cikm/MaureraDSSC20}. 
        }
        \label{fig:cw_heatmap}
    \end{minipage}
\end{figure*}

Video on demand and music streaming services are among the most successful application domains of Recommender Systems. 
Often, in a video-on-demand service (\eg Netflix, Amazon Prime Video) or on a music streaming platform (\eg Spotify) the user is provided with multiple rows of recommendations, each generated according to a specific criterion, \eg most recent, most popular, editorially curated (see Figure \ref{fig:amazon_netflix}). These rows are referred to as \emph{widgets}, \emph{shelves} or as \emph{carousels}.
In this scenario, the user satisfaction and behavior does not depend on a single recommendation list but rather on the entire set of recommendations provided in the various carousels, as well as their position. Finding appropriate combinations of algorithms and ranking them to provide the user with a personalized page is an active research topic of significant industrial interest  \cite{DBLP:conf/recsys/BendadaSB20,DBLP:conf/kdd/DingGV19,DBLP:conf/recsys/WuASB16}.

Despite this, in the traditional offline evaluation scenario each recommendation model is evaluated independently and the one with the highest quality is preferred. This evaluation procedure does not take into account how would the different recommendation lists complement each other in a real carousel user interface. As a consequence, it may lead to the selection of algorithms that provide similar sets of recommendations.
Since it is known that a set of diverse recommendations improves user satisfaction \cite{chen2018fast}
, and recommending the same item in multiple lists has little use, in some cases it will be beneficial to include algorithms with a lower individual recommendation quality if they generate recommendations with a different perspective.
Most articles targeting recommendations in a carousel setting are evaluated online with users of a certain platform. This puts a high resource requirement on researchers which will limit their ability to investigate this scenario. 
Furthermore, 
there seems to lack a standardized evaluation protocol to allow for offline experiments in a carousel setting. This is true, in particular, for how the two-dimensional structure of the user interface is taken into account in ranking metrics for which, to the best of our knowledge, no offline metric exists.

To address the highlighted issues, in this paper we propose a novel offline evaluation protocol that closely mirrors a real user interface with multiple carousels. In this setting, the recommendation quality of each model is computed by how much it improves the accuracy over one or more fixed carousels, in order to better evaluate the user satisfaction in such scenario. The contributions of this paper are as follows:
\begin{itemize}
    \item We propose an offline evaluation protocol based on real industrial carousel settings and provide experimental results highlighting the different relative accuracy of models evaluated in this way;
    \item We extend the widely used NDCG metric \cite{DBLP:conf/sigir/JarvelinK00}
    to a two-dimensional layout, that takes into account the user exploration behavior while navigating carousels. To the best of our knowledge, there exists no ranking metric that takes into consideration a two-dimensional layout.
\end{itemize}

The rest of the paper is organized as follows. Section \ref{sec:related_works} presents related works on carousel interfaces. Section \ref{sec:evaluation_protocol} provides a description of the carousel scenario and
the evaluation protocol.
Section \ref{sec:experimental_anaysys} reports the results of our experimental analysis. Finally, Section \ref{sec:conclusions} draws conclusions and presents possible future works.




\section{Related Work}
\label{sec:related_works}
Most of the works targeting a carousel user interface come from industrial research. 
This demonstrates the importance, at the industrial level, of identifying an adequate combination of carousels to show to the user.
However, it may also indicate that the lack of a standardized offline evaluation protocol is hampering researchers who do not have easy access to online evaluation infrastructure, preventing them to work on the topic.


\citet{DBLP:conf/recsys/WuASB16} analyze the problem of optimizing the position of the carousels on the interface of Netflix, a popular online video streaming service. 
The authors propose a graphical model based on the notion that the benefit of recommending a certain item depends on how novel it is relatively to the previous recommendations, a concept referred to as \textit{submodularity}. Their algorithm is able to leverage scrolls and navigation feedback to dynamically optimize the user interface. The evaluation is done both online and offline. For the offline evaluation standard metrics are reported (MRR and Precision) considering a carousel as a single item which is relevant if it contains at least a relevant recommendation, therefore not accounting for the ranking within the carousel. 
\citet{DBLP:conf/wsdm/GrusonCCMHTC19} optimize the homepage of Spotify, an online music streaming service.
The paper evaluates a series of policies to rank the most relevant playlists for a user. The policies are ranked according to first an online evaluation and then an offline evaluation. The two rankings are then compared and de-biasing techniques are discussed to improve the correspondence of offline evaluations with online ones.
In the offline evaluation, the carousels are sequentially concatenated as a single long recommendation list. This evaluation procedure, however, does not take into account the behavior of a user while exploring a two-dimensional interface. Again, the article points out at discrepancies between the online and offline evaluations.
\citet{DBLP:conf/recsys/BendadaSB20} propose a contextual multi-armed bandit online approach to optimize the two-dimensional layout of the homepage of an online music streaming service, called Deezer. Each user is shown a set of carousels which the user can swipe to reveal further items but the system does not log all user actions. To estimate which items the user has seen in the hidden part of the carousels they rely on the \textit{cascade model} \cite{DBLP:conf/wsdm/CraswellZTR08} assuming the user has seen all items before the one they interacted with and therefore has swiped and ignored them. The policies are evaluated both online and offline with expected cumulative regrets.
\citet{DBLP:conf/kdd/DingGV19} targets the problem of whole page optimization for the homepage of Amazon Video, a video streaming service. 
They assume that a set of carousels is already available and that the objective is to select which carousels to show and in which order while also accounting for the business constraints of the homepage. 
It should be noted that the personalization of homepage is a widely discussed problem also in other domains. For example,
\citet{DBLP:conf/www/AgarwalCYZ15} propose an optimization framework for the personalized widget layout on Linkedin, a known professional social network site, to improve user engagement. As opposed to the video-on-demand or music streaming scenario where there are many carousels available, in \cite{DBLP:conf/www/AgarwalCYZ15} there is only one.
Finally, some works try to account for the two-dimensional structure of the page during the algorithm training phase. For example, \citet{DBLP:conf/recsys/ElahiC20} propose to model the user response to the two-dimensional interface as an embedding.

\section{Carousel Evaluation Characteristics}
\label{sec:evaluation_protocol}
In this section we describe the characteristics of a real user interface in an industrial setting using a carousel layout.
\begin{description}
    \item[Interface:] A two-dimensional user interface (\idest a grid of recommendations) composed of different carousels (\idest rows). 
    In the general case, carousels may have different lengths which we assume remain constant for all users.
    
    \item[User Behaviour:] The user will not explore the interface one carousel at a time, in a sequential way. Rather, they tend to explore a two-dimensional carousel interface focusing on the top-left corner 
    \cite{DBLP:conf/recsys/ZhaoCHK16}
    (see also Figure \ref{fig:cw_heatmap}). 
    
    \item[Recommendations:] Each carousel is generated with a different algorithm or may come from a different provider, which may know the general layout of the page and the number of carousels, but will be unaware of the specific recommendations contained in the other carousels\footnote{This is common, with \emph{content aggregators} that aggregate carousels from different providers. For instance, a video content aggregator may have a carousels from Sky, Youtube, Netflix, Prime Video, etc.}. In general, no post-processing step is applied. Hence, the same item may be recommended multiple times across different carousels\footnote{For example, in the Netflix homepage shown in Figure \ref{fig:amazon_netflix} the TV series \emph{Space Force} appears both in the \emph{TV Comedies} and \emph{New Releases} carousels.}.

\end{description}

\subsection{Evaluation metrics}
\label{sec:evaluation_metrics}
Evaluation in a carousel setting presents broad similarities with a traditional top-n recommendation scenario. An important difference is the presence of duplicates in the recommendation list and the two-dimensional way users explore the interface, which impacts how ranking metrics may consider the item position. 
In a carousel evaluation scenario, an item in the recommendation list is relevant, \idest a correct recommendation, if it meets two conditions:
\begin{itemize}
    \item The item appears in that user's ground truth
    \item The item has been recommended only once or, if it has duplicates, it is the one corresponding to the \emph{best ranking}. 
    For traditional top-n ranking with a single recommendation list, duplicates are removed from the list; therefore, we assume that each carousel does not contain duplicates. However, duplicates might occur between carousels. Such duplicates must not be removed during the evaluation, in order to mimic the real behavior of carousel-based user interfaces.
    Accuracy metrics are not sensitive to the ranking of items; as such, their measurement with carousel-based used interfaces is the same as with traditional single-list recommender systems.

    On the other hand, in order to account for the two-dimensional user exploration of the interface a different definition of ranking discount is needed, as described in Section \ref{sec:two_dim_NDCG}.
\end{itemize}


\subsubsection{Two-dimensional NDCG}
\label{sec:two_dim_NDCG}
The idea of defining a two-dimensional ranking metric stems from the observation that users do not explore each carousel sequentially, but rather start from the top-left corner of the screen and proceed to explore the recommendations both to the right and to the bottom. This behavior has been known for many years in the Information Retrieval field and has been widely researched \cite{DBLP:conf/wsdm/ChierichettiKR11}. 
Understanding how the user attention varies with more complex interfaces is an active research field but is beyond the scope of this paper. The same phenomenon can be observed for recommender systems with a carousel interface. Figure \ref{fig:cw_heatmap} shows the number of interactions for items displayed in the carousel interface of a video-on-demand service, we can again see how the user interactions are concentrated in the  top-left corner \cite{DBLP:conf/cikm/MaureraDSSC20}.
To the best of our knowledge, no ranking metric that takes this behavior into account exists. As a way to approximate this behavior, we propose to extend the commonly used NDCG to the two-dimensional interface. We call this metric \emph{NDCG2D}, which will weight the item position approximating the two-dimensional position bias.

Traditional NDCG is defined as $NDCG = DCG/IDCG$ where DCG is the \textit{Discounted Cumulative Gain} computed from the ranking of the relevant items in the recommendation list and IDCG is the \textit{Ideal Discounted Cumulative Gain} defined as the DCG of the ideal ranked list, \idest the list composed of all the user's relevant items that can fit in the recommendation list length, ranked according to the ground truth relevance. Given $c$ as the cutoff, \idest recommendation list length, $rel(i)$ as the relevance of the item in position $i$, DCG can be computed as shown in Eq. \ref{eq:dcg}.

\begin{figure*}
    \begin{minipage}{0.45\textwidth}
        \begin{equation}
        \label{eq:dcg}
            DCG = \sum_{i=1}^{c} \frac{2^{rel(i)} - 1}{log_2(i+1)}     
        \end{equation}
    \end{minipage}
    \begin{minipage}{0.5\textwidth}
        \begin{equation}
        \label{eq:dcg2d}
            DCG2D = \sum_{i=1}^{l} \sum_{j=1}^{c}  \frac{2^{rel(i,j)} - 1}{log_2(\alpha i+ \beta j)}
        \end{equation}
    \end{minipage}
\end{figure*}


To account for the two-dimensional position bias, we define as $l$ the number of recommendation lists (\idest carousels) and extend the relevance function to two dimensions as $rel(i,j)$. The item relevance will be discounted by a quantity proportional to its position in both dimensions.
Thus, we define DCG2D as shown in Eq. \ref{eq:dcg2d}.
This metric
allows to give importance to both dimensions, according to the weights ($\alpha, \beta \geq 1$) provided, which can vary depending on the use case. 
Accordingly, we define its normalized version as $NDCG2D = DCG2D/IDCG2D$. Similarly to NDCG, the IDCG2D will be the DCG2D of the \textit{ideal ranking}, which is the matrix composed of the user's 
most relevant items, ranked according to the previously defined two-dimensional position discount, following constraint:
for any pair of cells $(i,j), (h,k)$ of the matrix,  $ rel(i,j) \geq rel(h,k)  $ if $ \alpha i+ \beta j < \alpha h+ \beta k $.
Notice how, in the case of different carousel lengths, the NDCG2D metric can be easily computed by assuming that all the carousels have the maximum carousel length and by treating the missing recommendations at the end of the shorter carousels as simply non-relevant.

\section{Experimental Analysis}
\label{sec:experimental_anaysys}
In this section we apply the proposed carousel evaluation on 
widely used algorithms and compare the results obtained with the traditional evaluation which considers each model independently. We discuss the results of this comparison and highlight some common trends and differences. We release the source code for our experiments in an online repository.\footnote{\url{https://github.com/nicolo-felicioni/RecSysCarouselEvaluation}}

\subsection{Algorithms} 
In our evaluation, we included several algorithms developed in the last three decades of research, trying to obtain a broad picture of different families of models.

\begin{itemize}
    \item \emph{Non-Personalized.} A simple but effective 
    model, \textbf{TopPopular} recommends to all users the most popular items. 
 
    \item \emph{Nearest-Neighbor Methods.} 
    We include in our analysis collaborative filtering (CF) nearest-neighbor techniques 
    such as \textbf{ItemKNN} 
    , based on item-item similarities and \textbf{UserKNN} 
    based on user-user similarities. In both cases the similarity is computed with cosine similarity with shrinkage.
    
    \item \emph{Graph-based Methods.} 
    We select two simple approaches that model a random walk in a graph containing user and item nodes.
    In \palphabold \cite{DBLP:conf/www/CooperLRS14} the similarity between items is computed as the transition probability between them. 
    \pbetabold  \cite{DBLP:journals/tiis/PaudelCNB17} extends \palpha dividing the similarity between two items by their popularity, raised to the power of an additional hyperparameter $\beta$, in order to reduce the popularity bias of the algorithm.

    \item \emph{Content-based and Hybrid Methods.} To account for content information we also include content-based models. Among the simplest content based models are neighborhood-based methods that build item-item similarities based on features, \textbf{ItemKNN-CBF} computes the item-based similarity using the item features while \textbf{UserKNN-CBF} computes the user-based similarity using user features. We use the cosine similarity with shrinkage.
    These methods can be easily extended by creating a new feature vector which concatenates both collaborative and content data 
    . The resulting hybrid algorithms are \textbf{ItemKNN-CFCBF} and \textbf{UserKNN-CFCBF}.
    
    \item \emph{Machine Learning Approaches.} We include several simple but well-known models relying on machine learning, like \textbf{SLIM ElasticNet (EN)} \cite{levy2013SLIM_ElasticNet}, a scalable variant of the original SLIM
    , and \textbf{SLIM BPR}, a variant of SLIM minimizing the BPR loss. 
    We also include a recent method called \textbf{\EASER} \cite{DBLP:conf/www/Steck19}, where the author showed how an "embarrassingly shallow" linear model with closed-form solution can outperform much more complex techniques. 
    
    \item \emph{Matrix Factorization Techniques.} We include various matrix factorization algorithms, like \textbf{PureSVD} \cite{DBLP:conf/recsys/CremonesiKT10}, \textbf{FunkSVD}
    , which are developed for explicit ratings, and \textbf{iALS} 
    , \textbf{Non-negative Matrix Factorization (NMF)} 
    , \textbf{Matrix Factorization (MF) BPR} 
    , focusing on implicit feedback.

\end{itemize}

\subsection{Hyperparameter optimization} 
While in this paper we do not aim to show that any particular model is superior to others, we nonetheless ensure that all algorithms are 
consistently optimized. To do so we followed the best practices highlighted by \citet{10.1145/3434185} and we relied on the framework published, using a Bayesian search \cite{DBLP:conf/recsys/FelicioniDCBHBD20} 
optimizing the MAP metric at cutoff 10.

\subsection{Datasets}
We report the results for some widely used  publicly available datasets.
We only selected datasets from domains that tend to use the carousel user interface, in particular video-on-demand. 
We included \textit{MovieLens 10M}, 
, a popular dataset of movies recommendations, with 69,878 users, 10,681 items and 10M ratings. The dataset contains user provided tags for items as well as the year of release and the genre.
We conducted experiments also on the \textit{Netflix} dataset 
, which is the well known movie dataset from the \textit{Netflix Prize}. To reduce the computational time, we randomly sampled 20\% of the users. After pre-processing, this dataset has 95,965 users, 17,768 items and almost 20M ratings. 
After pre-processing, we randomly selected 80\% of interactions for the training set and 10\% for both validation and test set.

\begin{table*}[]
    \centering
    \resizebox{0.94\textwidth}{!}{%
\begin{tabular}{l|ccc|cccc|cc|cc|c|}
\toprule
    {} & \multicolumn{3}{c|}{Individual} & \multicolumn{4}{c|}{Carousel (SLIM EN)} &  \multicolumn{2}{c|}{Improvement on SLIM EN} &       \multicolumn{3}{c|}{MAP rank}      \\
{} &      PREC &    MAP &   NDCG &     PREC &    MAP &   NDCG & NDCG 2D &     Individual &           Carousel & Individual & Carousel & $ \Delta $ rank \\
\midrule
SLIM EN     &         0.2460 & 0.2340 & 0.2856 &        -- & -- & -- &  -- &                           -- &                           -- &              -- &            -- &             -- \\
\midrule
\midrule
TopPop              &         0.0975 & 0.0709 & 0.0983 &        0.1399 & 0.1895 & 0.2967 &  0.2939 &                         -69.7\% &                          +4.8\% &             13 &            13 &              0 \\
\midrule
UserKNN CF          &         0.2343 & 0.2251 & 0.2815 &        0.1528 & 0.1955 & 0.3137 &  0.3225 &                          -3.8\% &                          +8.1\% &              1 &             3 &              -2 \\
ItemKNN CF          &         0.1885 & 0.1728 & 0.2122 &        0.1455 & 0.1921 & 0.3015 &  0.3034 &                         -26.2\% &                          +6.3\% &              8 &             9 &               -1 \\
\palpha             &         0.1646 & 0.1414 & 0.1915 &        0.1433 & 0.1912 & 0.3009 &  0.3021 &                         -39.6\% &                          +5.7\% &             12 &            10 &              +2 \\
\pbeta             &         0.1886 & 0.1686 & 0.2160 &        0.1430 & 0.1908 & 0.3014 &  0.3026 &                         -28.0\% &                          +5.5\% &             9 &            11 &              -2 \\
\midrule
\EASER              &         0.2260 & 0.2070 & 0.2566 &        0.1430 & 0.1899 & 0.3017 &  0.3012 &                         -11.5\% &                          +5.1\% &              4 &            12 &              -8 \\
SLIM BPR            &         0.2274 & 0.2159 & 0.2699 &        0.1490 & 0.1937 & 0.3084 &  0.3138 &                          -7.7\% &                          +7.2\% &              2 &             6 &              -4 \\
\midrule
MF BPR              &         0.1759 & 0.1502 & 0.1882 &        0.1479 & 0.1937 & 0.3040 &  0.3069 &                         -35.8\% &                          +7.2\% &             11 &             5 &              +6 \\
MF FunkSVD          &         0.2039 & 0.1748 & 0.2307 &        0.1560 & 0.1979 & 0.3148 &  0.3248 &                         -25.3\% &                          +9.5\% &              7 &             2 &              +5 \\
PureSVD             &         0.2217 & 0.2060 & 0.2527 &        0.1471 & 0.1924 & 0.3039 &  0.3061 &                         -12.0\% &                          +6.4\% &              5 &             7 &              -2 \\
NMF                 &         0.1872 & 0.1613 & 0.1974 &        0.1484 & 0.1938 & 0.3037 &  0.3064 &                         -31.1\% &                          +7.2\% &             10 &             4 &              +6 \\
IALS                &         0.2329 & 0.2152 & 0.2539 &        0.1592 & 0.1998 & 0.3101 &  0.3174 &                          -8.1\% &                         +10.5\% &              3 &             1 &              +2 \\
\midrule
ItemKNN CBF         &         0.0113 & 0.0052 & 0.0079 &        0.1264 & 0.1826 & 0.2875 &  0.2765 &                         -97.8\% &                          +1.0\% &             14 &            14 &              0 \\
\midrule
ItemKNN CFCBF       &         0.1952 & 0.1790 & 0.2174 &        0.1460 & 0.1923 & 0.3021 &  0.3044 &                         -23.5\% &                          +6.4\% &              6 &             8 &              -2 \\
\bottomrule
\end{tabular}
    }

    \resizebox{0.4\textwidth}{!}{%
    \centering
    \includegraphics[width=\columnwidth]{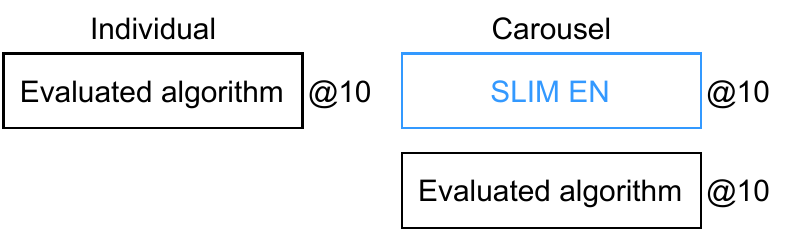}
    }
    
    \caption{Comparison of the MovieLens10M accuracy metrics with individual and carousel evaluation (with SLIM EN fixed as the first carousel) at recommendation list length of 10. Note that in the carousel evaluation there will be two recommendation lists. The improvement over the SLIM EN carousel is computed with MAP.}
    \label{tab:ml10macc_slim}
\end{table*}

\subsection{Discussion on the results}


For each algorithm, we report both its individual recommendation quality and the evaluation under a carousel setting when the first carousel is fixed and each algorithm is used to fill the second carousel (see Table \ref{tab:ml10macc_slim}). All recommendation lists have a length, \idest cutoff, of 10, however note that in the individual evaluation there will be a single recommendation list while in the carousel evaluation there will be more than one, in this case two. We evaluated two cases, in the first one the fixed carousel is a non-personalized TopPopular model, while in the second it is SLIM EN, the personalized model with the highest individual MAP. 
For the NDCG2D, we chose as an example $\alpha=\beta=1$ in which the user explores both horizontally and vertically with equivalent penalty. 
The results obtained with the TopPop and SLIM EN fixed carousels and for both datasets are broadly similar. For space reasons we will only discuss the results for MovieLens10M with a SLIM EN fixed carousel (see Table \ref{tab:ml10macc_slim}). We provide the full results in the online material.

Based on the MAP of the individual evaluation, the best performing algorithms are SLIM, UserKNN and IALS. 
However, when looking at the results of the carousel evaluation, we can see that many differences emerge. For example, \palpha has an individual MAP that is 39.5\% lower than SLIM EN, while \EASER only 11.5\% lower. In the carousel evaluation instead, the improvement provided by these two algorithms is similar with \palpha being slightly better. The discrepancy between individual and carousel evaluation is even clearer if we look at the column showing the difference in the rankings of all algorithms in the two scenarios. 
As a general trend we can see that the relative performance of the models differ, resulting in several changes to the ranking of the algorithms. Some models, in this case all matrix factorization algorithms (except for PureSVD) gain several positions. NMF and MF BPR jump up by 6 positions. On the other hand, item-based machine learning models tend to lose some positions. 
As a result, in the carousel evaluation the best algorithms are IALS, FunkSVD and UserKNN.
The difference in those rankings lies in how those recommendations intersect. Algorithms which will tend to recommend items similar to the ones provided by SLIM EN will be penalized in this carousel evaluation, whereas algorithms providing accurate but different recommendations
will be advantaged. An interesting case is \EASER, which loses 8 positions, probably because of its resemblance with the SLIM algorithm.

Regarding the NDCG2D metric, we can notice how it too can lead to different decisions with respect to the standard NDCG in the carousel evaluation, though to a lesser extent.
Looking at the results in Table \ref{tab:ml10macc_slim}, we notice that the NDCG results of \EASER and Item KNN are very similar under the carousel evaluation, but NDCG2D will give a slight preference for Item KNN instead. This difference will become more marked as the page layout becomes more complex including more and longer carousels.

\section{Conclusions and Future Work}
\label{sec:conclusions}
This paper proposes a new offline evaluation protocol for a carousel user interface, where the recommendation quality of a model is not measured independently but rather is put into the context of other recommendation lists being already available to the users. The experimental analysis shows that the relative ranking of the personalized algorithms changes when a carousel is fixed as the first displayed to the user.
This is in line with previous observations that the correlations between models have an important role to play and should be taken into account during offline evaluation as well. Further study is needed to better understand this impact. 
Important future work is an online study
to measure how closely the offline carousel evaluation is able to represent the user behavior, as well as extended offline evaluation, involving a wider array of scenarios, for example sequential recommendation and also including families of models not analyzed in this paper.

Ultimately, the carousel evaluation protocol will allow researchers to conduct offline evaluations in these industrially relevant scenarios and open a wide number of research possibilities in studying how to combine the strength of various models and techniques to provide the user with ever more accurate and interesting recommendations.



\bibliographystyle{ACM-Reference-Format}
\bibliography{main.bib}

\end{document}